\documentclass[10pt,prd,onecolumn,nofootinbib,
eqsecnum,byrevtex,amsfonts,amssymb,amsmath]{revtex4}
\usepackage[dvips]{graphicx}
\usepackage[scriptsize,bf]{caption}
\usepackage{xspace}
\newsavebox{\oldtext}

\newcommand{\eps}{\epsilon}
\newcommand{\br}[1]{\left(#1\right)}
\newcommand{\ud}[1]{\mathrm{d}#1}
\renewcommand{\exp}[1]{e^{#1}}
\newcommand{\sq}[1]{\left[#1\right]}
\newcommand{\olarge}[1]{O\left(#1\right)}
\newcommand{\osmall}[1]{o\left(#1\right)}

\newcommand{\abs}[1]{\left|#1\right|}
\newcommand{\tr}[1]{\mathrm{Tr}\left({#1}\right)}
\newcommand{\tc}{$\mathbb{R}\!\times\!\mathbb{S}^2$\xspace}
\newcommand{\ts}{$\mathbb{S}^3$\xspace}
\newcommand{\Exp}[1]{\mathrm{exp}\br{#1}}
\newcommand{\ha}[1]{$\mathbf{H}_{#1}$}
\newcommand{\sk}[1]{$\mathbf{S}_{#1}$}
\newcommand{\su}{$\mathrm{SU}(2)$\xspace}
\begin{document}
\title{Skyrmion on a three--cylinder}
\author{{\L}ukasz \surname{Bratek} }
\email[]{Lukasz.Bratek@ifj.edu.pl}
\affiliation{The Henryk Niewodnicza{\'n}ski Institute of Nuclear Physics, \\
Polish Academy of Science, \\ Radzikowskego 152, 31-342
Krak{\'o}w, Poland.}
\date{v2 1Apr2008}

\begin{abstract}
   The class of static, spherically symmetric, and finite energy hedgehog solutions
 in the \su Skyrme model is examined on a metric three-cylinder. The exact analytic shape function of
 the 1-Skyrmion is found. It can be expressed via elliptic integrals. Its
energy is calculated, and its stability with respect to radial and
spherically symmetric deformations is analyzed. No other
topologically nontrivial solutions belonging to this class are
possible on the three-cylinder.
\end{abstract}
 \maketitle

\section{Introduction and motivation}\label{sec:intro}
We discuss static and spherically symmetric "hedgehog" fields in
the \su Skyrme model on a metric three-cylinder
(\tc).\footnote{Throughout this paper we distinguish the notion of
'metric three-cylinder' from that of 'topological three-cylinder'.
We use \tc as shorthand for 'metric three-cylinder' with line
element $\ud{\psi^2}+\ud{\theta}^2+\sin^2{\theta}\ud{\phi}^2$. The
same remark applies to a three-sphere
--- \ts is shorthand used in this paper for 'metric three-sphere' with standard line element
$\ud{\chi^2}+\sin^2{\chi}\br{\ud{\theta}^2+\sin^2{\theta}\ud{\phi}^2}$}
A static and spherically symmetric spacetime line element with \tc
as hypersurfaces of constant time reads
$$\ud{s}^2=-\,\ud{t}^2+L^2\br{\ud{\psi}^{2}+\ud{\Omega}^2\br{\theta,\phi}},
\qquad
\ud{\Omega}^2\br{\theta,\phi}=\ud{\theta}^{\,2}+\sin^2{\theta}\,\ud{\phi}^{\,2}.$$
Coordinates $\theta$ and $\phi$ are standard spherical angles and
$\psi\in(-\infty,+\infty)$.\footnote{More generally, we could also
set $-f^2(\psi)\ud{t}^2$ with any $f$ in the line element, but the
rate of clocks is irrelevant for static field configurations since
the kinetic term of such fields vanishes.} The positive scale
factor $L$ can be interpreted as the three-cylinder's radius, but
its role here is to provide the Skyrme model with additional
length scale that can be compared with the characteristic soliton
size. This simple field-theoretical setup leads to a single
equation for the Skyrme hedgehog that can be solved exactly in
three spatial dimensions. The analogous solution on the metric
three-sphere (\ts) can be found only by numerical integration.

By considering fields with appropriate asymptotics, \tc (with
attached two "points at infinity") may be topologically identified
with \ts. In contrast to the \ts case, the metric geometry of \tc
is not isotropic -- the sectional Gauss curvature of \tc is
direction dependent. Nevertheless, these two geometries are
conformally identical:
$$\ud{\psi^2}+\ud{\Omega}^2\br{\theta,\phi}=\frac{\ud{\chi}^2+\sin^2{\chi}\,\ud{\Omega}^2\br{\theta,\phi}}{\sin^2{\chi}},
\qquad \chi=2\arctan{\br{\exp{\psi}}}\in(0,\pi)$$ (length scales
on \tc and on \ts are assumed equal). Coordinate $\chi$ is the
standard third
 angle on \ts. This conformal identification of metric
geometries of \tc and \ts allows for identification of the
translation symmetry Killing vector $\partial_{\psi}$ on \tc with
the conformal symmetry Killing vector
$\sin\!{\chi}\partial_{\chi}$ on \ts.

The conformal symmetry Killing vector is related to bifurcations
of static spherically symmetric hedgehogs on \ts at characteristic
critical length scales. In particular, at the critical radius
$L_c=\sqrt{2}$, the $1$-Skyrmion (\sk{1}) on \ts separates from
the identity map
\ha{1}$:\mathbb{S}^3\to\mathrm{SU}(2)\sim\mathbb{S}^3$ by a
conformal deformation of this map \cite{bib:mantsamol}, namely, in
the limit $L\searrow{}L_c$, the shape function of \sk{1} reads
$\chi\to\exp{\eps\sin{\chi}\partial_{\chi}}\chi=\chi+\eps\sin{\chi}+o(\eps)$,
$\eps\propto{\sqrt{L-L_c}}$.\footnote{Energy and shape function of
\sk{1} can be found with arbitrary accuracy by perturbations
(however, the resulting series seem to have finite radius of
convergence in $L-L_c$, presumably bounded by $\sqrt{2}$,
therefore one cannot reconstruct the energy of flat space
1-Skyrmion by taking the limit $L\to\infty$). To the leading order
this perturbative calculation confirms the discussed above
conformal deformation of the identity solution
\cite{bib:moriond}.} Note that $\sin{\chi}$ is also the
eigenfunction corresponding to the lowest eigenvalue of the
Hessian evaluated at \ha{1} \cite{bib:wirzb}. The eigenvalue
vanishes at $L=L_c$ and is negative for $L>L_c$ and positive for
$L<L_c$. The conformal deformation of \ha{1} generated by
$\sin{\chi}\partial_{\chi}$ is thus energetically preferable for
$L>L_c$. For such $L>L_c$ it is the \sk{1} which minimizes the
energy functional.

On \tc the situation is qualitatively different. A deformation of
a solution generated by $\partial_{\psi}$ is simply a translation
which is also a symmetry of the space. This deformation leaves
energy of the 1-Skyrmion on \tc  and the boundary conditions at
infinity unchanged. For that reason one should not expect
bifurcations similar to that on \ts to occur on \tc. Moreover, by
contrast with the \ts case, one may also expect that the
1-Skyrmion on \tc should uniformly tend to a harmonic map of the
related sigma model on \tc.

\medskip

Translation symmetry of Skyrme equations on \tc enables one to use
the conserved current associated with this symmetry  as the first
integral of these equations. In particular, it can be shown that
static, spherically symmetric, finite energy and topologically
nontrivial hedgehog solutions with topological charge other than
$\pm1$ cannot exist on \tc. One can also easily examine
properties of the 1-Skyrmion, calculate its energy, and analyze its
stability. Remarkably, all this can be done without knowing the
exact form of the 1-Skyrmion's shape function. Next, we construct
an approximate formula for the 1-Skyrmion's shape function, and
finally we find its exact form.

\section{Basic equations} The standardized
form of Lagrangian density of \su-valued Skyrme field $U$ in
spacetime with metric signature $(-,+,+,+)$ is \cite{bib:ati}
$$\mathcal{L}[U]=\sqrt{-g}\br{\frac{1}{2}\,\tr{K_{\mu}K^{\mu}}+
\frac{1}{16}\,\tr{\sq{K_{\mu},K_{\nu}}\sq{
K^{\mu},K^{\nu}}}},\qquad K_{\mu}\equiv{}U^{-1}\partial_{\mu}U.$$
This Lagrangian simply generalizes to other matrix-valued fields.
In the construction of this density, the principle of minimal
coupling of matter with gravitation is assumed -- metric tensor
couples with matter fields in the same way it does in Minkowski
spacetime with arbitrary
 coordinate system.  The Skyrme field behaves
as a scalar with respect to spacetime transformations. The first
summand in the Lagrangian is known as the sigma term. The second
term has the opposite scaling with the length scale. It was
introduced by Skyrme \cite{bib:skyr} to ensure the existence of
solitons among static solutions.

We use (metrical and topological) isomorphism of the \su group and
the unit three-sphere: $\mathbb{S}^3\ni\br{\Psi,\Theta,\Phi}\to{}
U=\mathrm{exp}\br{i\,\Psi\,\vec{\sigma}\circ\vec{n}(\Theta,\Phi)}\in{}\mathrm{SU}(2)$,
where $\vec{n}(\Theta,\Phi)$ is a unit direction determined by
spherical angles $(\Theta,\Phi)$:
$\vec{n}=\sq{\sin{\Theta}\cos{\Phi},\sin{\Theta}\sin{\Phi},\cos{\Theta}}$,
${\vec{\sigma}}$ is a vector of Pauli matrices, and $\Psi$ is the
third angle on \ts.

As general relativity theory teaches us, the stress tensor of
matter fields is proportional to the variational derivative of the
action functional of these fields with respect to the spacetime
metric tensor
$$T_{\mu\nu}=-\frac{2}{\sqrt{-g}}\frac{\delta\,\mathcal{S}}{\delta\,{g^{\mu\nu}}},\qquad
\mathcal{S}[U]=\int\,\mathcal{L}[U].$$ For the Skyrme Lagrangian
$$T_{\mu\nu}=\frac{\mathcal{L}[U]}{\sqrt{-g}}\,g_{\mu\nu}
-\tr{K_{\mu}K_{\nu}}
-\frac{1}{4}\tr{\sq{K_{\mu},K_{\alpha}}\sq{K_{\nu},K^{\alpha}}}.$$
The energy functional of static {and} spherically symmetric Skyrme
hedgehogs on \tc,  reduces to\footnote{{If $\xi^{\mu}$ is a
Killing vector and if $T^{\mu}_{\phantom{\mu}\nu}$ is
divergence-free then the three-form
$\omega=-\frac{1}{3!}T^{\mu}_{\phantom{\mu}\nu}\xi^{\nu}\sqrt{-g}\epsilon_{\mu\alpha\beta\gamma}
\ud{x}^{\alpha}\wedge\ud{x}^{\beta}\wedge\ud{x}^{\gamma}$ is
closed, that is $\ud{\omega}=0$. In particular, integrated over
\tc with $\xi^{\mu}\equiv\br{\partial_t}^{\mu}$,  this form
defines energy functional (\ref{eq:enfunct}). Provided $\omega$
vanishes at spatial infinity sufficiently fast, this energy is
conserved since $\ud{\omega}=0$.}}
[$F'\equiv\frac{\ud{F(\psi)}}{\ud{\psi}}$]
\begin{equation}\label{eq:enfunct}
E[F]=4\pi\int\limits_{-\infty}^{+\infty}\ud{\psi}\sq{
L\br{2\sin^2{F}+F'^2}+\frac{1}{L}\sin^2{F}\br{\sin^2{F}+2F'^2}
}.\end{equation}  We have used the hedgehog ansatz $\Psi=F(\psi)$,
$\Theta\equiv\theta$, and $\Phi\equiv\phi$.

To ensure finiteness of energy $E[F]$ we impose the following
asymptotic (finite energy) condition:
{$\sin\br{F}=\osmall{|\psi|^{-1/2}}$} {as $\abs{\psi}\to\infty$}.
Under this condition {$U\to\pm\mathbf{1}$} at infinity. Solutions
with such asymptotics are characterized by the topological charge
$Q=\br{F(+\infty)-F(-\infty)}/\pi$ {and their} energies are
bounded from below by a positive number {$12\pi^2\abs{Q}$} [this
is known as the Faddeev--Bogomolny bound, which is universal for
the \su Skyrme model in three spatial dimensions; {cf.
\cite{bib:krush} for a proof}]. We employ this distinguished
$12\pi^2$ value as the unit of energy.

Critical points of energy functional (\ref{eq:enfunct}) are
solutions of the following equation:
\begin{equation}\label{eq:eq_sec_ord}
\br{1+\frac{2}{L^2}\sin^2{F}}F''+\frac{F'^2}{L^2}\sin{2F}-\br{1+\frac{1}{L^2}\sin^2{F}}\sin{2F}=0.\end{equation}
A Noether current associated with the Killing vector
$\partial_{\psi}$ of the translation symmetry of  \tc is
$j^{\mu}=T^{\mu}_{\phantom{\mu}\nu}\br{\partial_{\psi}}^{\nu}$.
{Since the Lagrangian is also $\psi$-translation invariant, this
current is conserved for solutions.} The conservation equation
$0=\nabla_{\mu}j^{\mu}\equiv\partial_{\mu}\br{\sqrt{-g}j^{\mu}}/\sqrt{-g}$
{yields on integration with respect to $\psi$} {
\begin{equation}\label{eq:current}\br{ {{L}}^2 + 2\,{\sin^2 F} \right) \,
      {F'}^2 - {\sin^2 F}\,\left( 2\,L^2 + {\sin^2 F}
      }=C,\end{equation}
}{which is} the first integral of Eq. (\ref{eq:eq_sec_ord}).
The finite energy {condition} implies the integration constant $C$
in (\ref{eq:current}) must be zero. Hence the first integral of
Eq. (\ref{eq:eq_sec_ord}) for a finite energy solution reads
\begin{equation}\label{eq:first_intg}F'^2=\frac{2L^2+\sin^2F}{L^2+2\sin^2F}\sin^2F.\end{equation}
Inversely, for nonconstant $F$, {on differentiating
(\ref{eq:current}) with respect to $\psi$,} we obtain
(\ref{eq:eq_sec_ord}).

Any nonconstant solution of (\ref{eq:eq_sec_ord}) {not satisfying}
(\ref{eq:first_intg}) must be either divergent or oscillate around
$\pi/2\,\br{\mathrm{mod}\,{\pi}}$, {depending on} the sign of $C$.
Indeed, if {$C>0$} at some point, then  $|F'|$ must be bounded
from zero for all $\psi$. Consequently, $F$ must be divergent as
$|\psi|$ tends to $\infty$. If {$C<0$} then (\ref{eq:current})
implies that $F'^2\leqslant2\sin^2F-\abs{C}L^{-2}$, and
(\ref{eq:eq_sec_ord}) implies that $F''=c(F)\sin{2F}$ for all
$\psi$, where $c(F)$ is a strictly positive function. As so, $F$
must {stay strictly within the strip
$(0,\pi)\,\br{\mathrm{mod}\,\pi}$} {oscillating} around
$\pi/2\,\br{\mathrm{mod}\,{\pi}}$. For such solutions energy
defined in (\ref{eq:enfunct}) cannot be finite.

\section{Properties of {static, spherically symmetric, and} finite
energy {hedgehog} solutions} As we have seen above, the only
solutions of Eq. (\ref{eq:eq_sec_ord}) that have finite
energy are solutions of Eq. (\ref{eq:first_intg}). Without
loss of generality we may assume that $0<F<\pi$ at some point.
Suppose that $F$ leaves the strip $[0,\pi]$ at another point.
Equation (\ref{eq:first_intg}) then implies that $F'$ vanishes at
this point. By uniqueness theorems applied to Eq.
(\ref{eq:eq_sec_ord}) the only solution of this initial problem is
the vacuum solution $\sin{F}\equiv0$. {Therefore, $0<F<\pi$} for
all finite  $\psi$.
 It also follows from (\ref{eq:first_intg}) that
$F'$ is bounded from zero within this strip, thus $F$ is monotonic
and attains $0$ {or} $\pi$ only asymptotically. Now it is clear
that this solution has unit topological charge. {It also follows}
that no other {static, spherically symmetric, and} finite energy
{hedgehog} solution with larger than unit topological charge can
exist on \tc.

Within the class of {static, spherically symmetric, and} finite
energy {hedgehog} solutions, we may restrict our attention to
considering only its representative for which $F(0)=\pi/2$ and
$F'(0)>0$ [then $F(-\infty)=0$ and {$F(+\infty)=\pi$}]. We shall
refer to the class of solutions as the 1-Skyrmion on metric
three-cylinder.

It follows from the first integral (\ref{eq:first_intg}) that
$$\frac{1}{2}\sin^2F<\frac{2L^2+1}{L^2+2}\sin^2F\leqslant{}F'^2\leqslant2\sin^2F.$$
This implies that the graph of the 1-Skyrmion's shape function,
passing through the point $F(0)=\pi/2$, is contained within the
region bounded by graphs of the limiting profiles
\begin{equation}\label{eq:limit_sol}2\arctan{\br{\exp{\psi/\sqrt{2}}}}<F<2\arctan{\br{\exp{\sqrt{2}\psi}}}.\end{equation}
The first is attained uniformly at $L\to0$, whereas the second one
at $L\to\infty$. This observation suggests that we can assume the
following test function
\begin{equation}\label{eq:approx}\widetilde{F}=2\arctan{\br{\exp{G(L)\psi}}}
\end{equation} to approximate the exact profile of the 1-Skyrmion. The optimum
value of $G(L)$  can be determined at a given $L$ by finding the
minimum energy of the test function. Hence $G(L)$ must satisfy the
condition $d_{G}E[\widetilde{F}]=0$, which gives
$$G(L)=\sqrt{\frac{2+6L^2}{4+3L^2}},\qquad
E[\widetilde{F}]=\frac{16\pi\sqrt{2}}{3L}\sqrt{\br{4+3L^2}\br{1+3L^2}}.$$
The energy {of $\widetilde{F}$ with this $G(L)$} attains its global
minimum $\frac{4\sqrt{6}}{3\pi}\times12\pi^2$ at $L={\sqrt{2/3}}$.

\section{Stability analysis of the 1-Skyrmion}
Let $\mathcal{H}$ denote the Hilbert space of test functions that
are {both} $\mathcal{C}^1$ {and normalizable to
$\br{4\pi{}L^3}^{-1}$} on $\mathbb{R}$.\footnote{Then
$\mathcal{H}$ extends to spherically symmetric functions
normalizable to unity on \tc (with volume element
$\ud{\mathcal{V}}=L^3\sin{\theta}\,\ud{\psi}\,\ud{\theta}\,\ud{\phi}$)}
Such functions must vanish at infinity {faster than}
$\abs{\psi}^{-1/2}$. Let $F(\psi)$ {be} the exact shape function
of the 1-Skyrmion {and $h\in\mathcal{H}$}. {Let substitute
$F_{\eps}(\psi)=F(\psi)+\eps\,h(\psi)$} to (\ref{eq:enfunct}) and
expand {the functional} as a power series in $\eps$; {then}
$$E[F+\eps\,h]=E[F]+\eps\,\delta_F{E}[F](h)+\eps^2\,\delta^2_FE[F](h,h')+\osmall{\eps^2}.$$
{The first variation} $\delta_F{E}[F](h)$ must vanish for any
$h\in\mathcal{H}$ on account of the Euler--Lagrange equations. The
Hessian {$\delta^2_FE[F](h,h')$} is {a} quadratic {form} in $h$
and $h'$ and {provides an energy measure}  of a perturbation $h$.
{To} find the lowest bound $\lambda_g$ of the Hessian attained
over $\mathcal{H}$ one has to find a conditional minimum of the
Hessian {with the condition that $\int{}h^2\ud{\mathcal{V}}=1$ and
with $\lambda_g$ being the corresponding Lagrange multiplier}. For
$h$ to be the minimum energy perturbation {in $\mathcal{H}$, it is
necessary that $h$ be the solution of the following eigenvalue
problem} \cite{bib:hilb}
$$\frac{\delta\br{\delta^2_FE[F](h,h')}}{\delta{}h(\psi)}=\lambda_g
\frac{\delta\br{\int{}h^2\ud{\mathcal{V}}}}{\delta{}h(\psi)}
=2\lambda_g{}h(\psi),\qquad h\in\mathcal{H}$$ {corresponding to
the lowest eigenvalue $\lambda_g$.} This {ordinary differential}
equation is linear in $h$, $h'$, and $h''$. {By substituting
$h\equiv{}F'$, it can be verified that the variational derivative
on the very left in the above equation vanishes identically,
provided $F$ satisfies (\ref{eq:first_intg}).}\footnote{{Actually,
$\delta^2_FE[F](h,h')$ vanishes for $h=F'$ by translation
invariance of functional (\ref{eq:enfunct}). For if
$F_{\eps}(\psi)=F(\psi+\eps)$, then
$E[F_{\eps}]=E[F+\eps{}F'+\olarge{\eps^2}]=E[F]+\eps\delta_F{}E[F]\br{F'+\olarge{\eps}}+
\eps^2\delta^2_FE[F]\br{F'+\olarge{\eps},F''+\olarge{\eps}}+\osmall{\eps^2}.$
But for solutions $\delta_F{}E[F]\equiv0$, hence, on expanding
once again,
$E[F_{\eps}]=E[F]+\eps^2\delta^2_FE[F](F',F'')+\osmall{\eps^2}$.
By translation invariance $E[F]={}E[F_{\eps}]$ for any $\eps$,
thus $\delta^2_FE[F](F',F'')$=0.}}
{Thus $F'$ is the eigenfunction of the Hessian corresponding to
the eigenvalue $\lambda_g=0$.}

We have seen that the 1-Skyrmion's shape function $F$ is monotonic
and that $F'=0$ only asymptotically. Thus $F'$ has no nodes for
finite $\psi$. It is known  that the eigenfunction corresponding
to the lowest eigenvalue has no internal nodes \cite{bib:hilb}. As
so, we arrive at the conclusion that the 1-Skyrmion on \tc is
(marginally) stable against radial and spherically symmetric
deformations.

\section{The exact analytic formula for the 1-Skyrmion on
\tc}\label{sec:exact} With the initial condition $F(0)=\pi/2$
{Eq.} (\ref{eq:first_intg}) can be rewritten as
$$\psi(F)=
\int\limits_{\pi/2}^{F}\frac{\ud{f}}{\sin{f}}\sqrt{\frac{L^2+2\sin^2f}{2L^2+\sin^2f}},
\qquad (0,\pi)\ni{}F\,\to\,\psi\in(-\infty,+\infty).$$ By
substituting $z=\sin{\br{\Phi(f)}}$, with $\Phi$ being related to
$f$ by
$$k\sqrt{2}\sin{\br{\Phi(f)}}=\sqrt{\frac{L^2+2\sin^2f}{2L^2+\sin^2f}},
\qquad k=\sqrt{\frac{2+L^2}{2+4L^2}},$$ this integral can be
brought into the form containing canonical elliptic integrals of
the first and of the third kind (see Appendix):
$$\abs{\psi(\Phi(F))}=\frac{3\sqrt{2}}{4\sqrt{1+2L^2}}\int\limits_{\sin{\br{\Phi(F)}}}^{1}\left( 1 + \frac{1}{1 - 4\,k^2\,z^2} \right) \,
  \frac{\ud{z}}{{\sqrt{\left( 1 - z^2 \right) \,\left( 1 - k^2\,z^2 \right)
  }}}\,.
$$
The definitions of $k$ and $\Phi$ are correct since
$1/2\leqslant{}k<1$ and $1/2<1/(2k)\leqslant\sin{\Phi}\leqslant1$.
We choose in this integral $\psi<0$ for $0<2F<\pi$, and $\psi>0$
for $\pi<2F<2\pi$. The resulting (although implicit)  exact
formula for the 1-Skyrmion on \tc reads
\begin{equation}\label{eq:exact_sol}\abs{\psi(F)}=\frac{3\sqrt{2}}{4\sqrt{1+2L^2}}
\br{\mathbf{F}\br{\frac{\pi}{2},k}-\mathbf{F}\br{\Phi(F),k}+
\mathbf{\Pi}\br{\Phi(F),-4k^2,k}-
\mathbf{\Pi}\br{\frac{\pi}{2},-4k^2,k}}.\end{equation}

\section{Energy of the 1-Skyrmion on \tc} The exact form of the
1-Skyrmion's shape function is not required to calculate {its}
total energy. By expressing $\ud{\psi}$ by $(F')^{-1}\ud{F}$ and
using the first integral (\ref{eq:first_intg}) in
(\ref{eq:enfunct}), we obtain the following formula for the energy
of the 1-Skyrmion:
$$E[F]=\frac{16\pi}{L}\int\limits_{0}^{\pi/2}\ud{F}\sin{F}\sqrt{\br{2L^2+\sin^2F}\br{L^2+2\sin^2F}}.$$
This integral can be equivalently rewritten as
$$E[F]=E(L)\equiv\frac{16\pi\sqrt{2}}{L}Q\sqrt{P}
\int\limits_0^{1/\sqrt{Q}}\ud{z}\sqrt{(1-z^2)(1-k^2z^2)},$$ where
we have substituted $\sqrt{Q}\,z(F)=\cos{F}$, $Q=1+L^2/2$,
$P=1+2L^2$, and $k=\sqrt{Q/P}<1$ ($k$ is the same as in Sec.
\ref{sec:exact}). This integral, in turn, can be expressed by
means of standard elliptic integrals of the first and second kind:
\begin{equation}\label{eq:en_exact}
E(L)=\frac{16\pi\sqrt{2}}{3L}\br{\sqrt{P}(P+Q)\mathbf{E}(k,\widetilde{\Phi})-
\sqrt{P}(P-Q)\mathbf{F}(k,\widetilde{\Phi})+\sqrt{(P-1)(Q-1)}}\end{equation}
where $\widetilde{\Phi}=\arcsin(1/\sqrt{Q})$.

Let us calculate the minimum energy of the 1-Skyrmion. The minimum
is attained at a radius $L_m$ defined by the equation $E'(L_m)=0$.
This equation cannot be solved by radicals. However, this equation
is analytic and can be expanded about some point. We have already
seen that the minimum radius should be close to $L=\sqrt{2/3}$.
Therefore, we substitute $L_m=\sqrt{2/3}+\zeta$ and find the
Taylor series expansion with respect to {the unknown and small}
$\zeta$. In the first order approximation we obtain
$$L_m=\frac{9\,{\sqrt{42}}\,\left( 11\,\mathbf{E}(\pi/3,2/\sqrt{7}) - 3\,\mathbf{F}(\pi/3,2/\sqrt{7}) \right)  +
    30\,{\sqrt{2}}}{{\sqrt{7}}\,\left( 409\,\mathbf{E}(\pi/3,2/\sqrt{7}) -
       141\,\mathbf{F}(\pi/3,2/\sqrt{7}) \right)  - 26\,{\sqrt{3}}}+o(\zeta)=0.81509\dots$$
To find a better approximation we now discard only the
$\mathcal{O}(\zeta^4)$ term in the expansion of
$E'(\sqrt{2/3}+\zeta)$, obtaining a cubic polynomial in $\zeta$.
Cardano's formulas for the roots of the cubic polynomial lead to a
monstrous expression for $L_m$. Therefore, here we show only the
decimal expansion of $L_m$ and the corresponding minimum of the
1-Skyrmion's energy:\footnote{{Note that $E(L_m)$ is very close to
the energy per baryon for cubic Skyrme crystal found numerically
by relaxations to be $\approx1.036\times12\pi^2$
\cite{bib:battyeSutcliffe} (no error of this value was given,
however). }}
\begin{equation}\label{eq:min_en}L_m=0.8150941506\cdots,\qquad
E(L_m)=1.03576803116479882348\cdots\times12\pi^2.\end{equation}
All of these digits {shown} are exact and follow from the
presented third order calculation. This statement may be verified
by carrying out a similar fourth order calculation (the roots of a
fourth order polynomial can be still found by radicals).

\section{Discussion}
\begin{figure}[h!]\centering
    \begin{tabular}{cc}
      \includegraphics[width=0.47\textwidth]{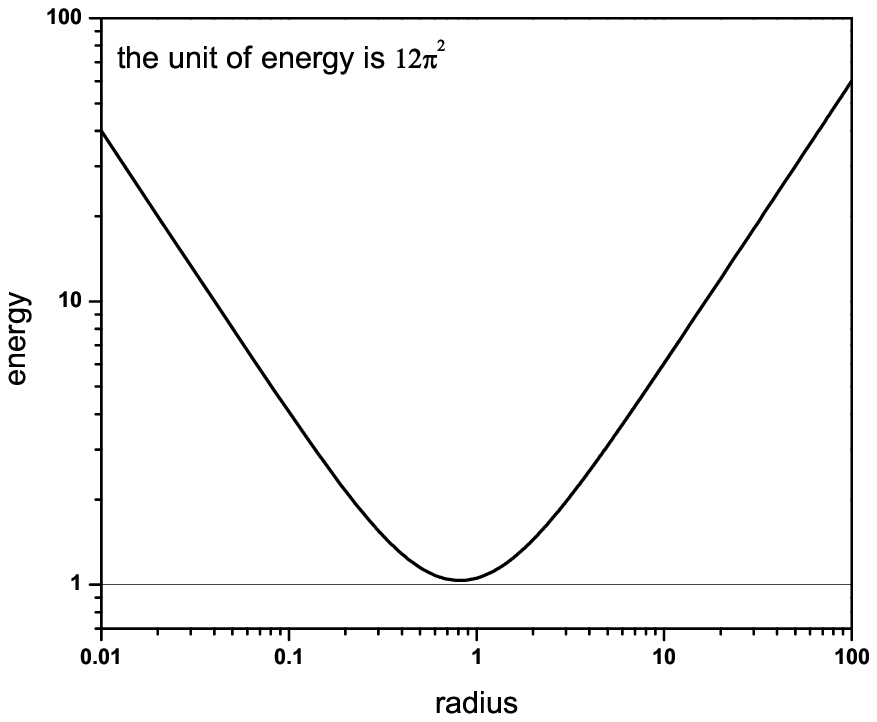}&
     \includegraphics[width=0.47\textwidth]{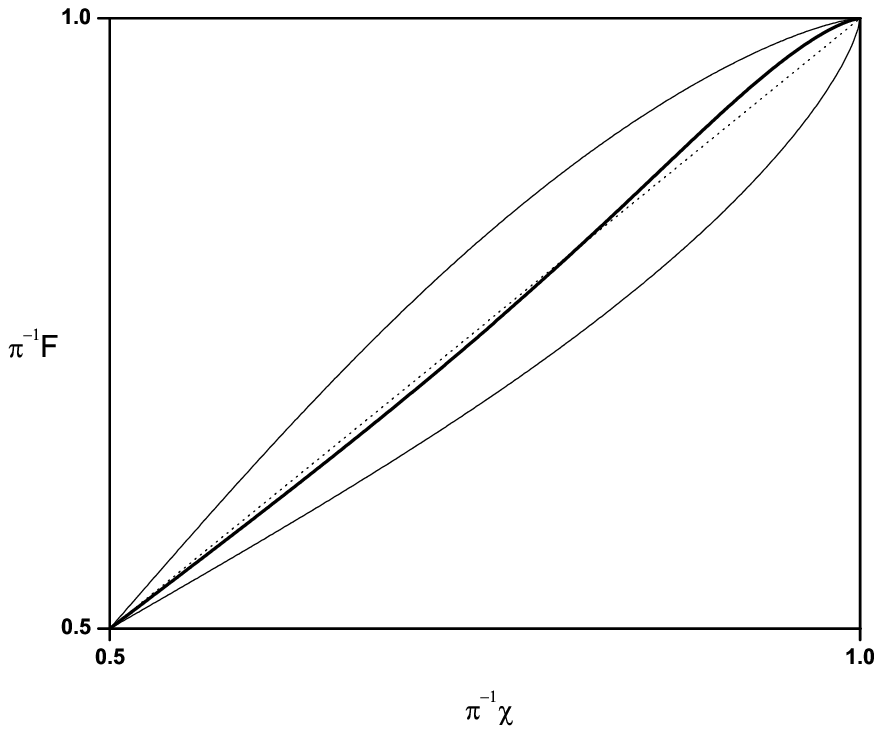}\\
     a)&   b)\\
         \end{tabular}
    \caption{\label{fig:}
    \ \textbf{(a)}\ Energy of the 1-Skyrmion on a metric three-cylinder shown as a function
    of the scale factor $L$. The
    unit of energy is $12\pi^2$. \ \textbf{(b)}\ Thick line -- the exact shape function (\ref{eq:exact_sol}) of the 1-Skyrmion
    with the minimum possible energy at $L=L_m$ [cf. Eq. (\ref{eq:min_en})]; dotted line -- the approximate shape
    function of the 1-Skyrmion defined in Eq. (\ref{eq:approx}) [with $L=\sqrt{2/3}$ and $G(L)=1$];
    thin lines -- the 1-Skyrmion in the limit $L\to0$ (lower line) and in the limit $L\to\infty$ (upper line).
    The $\chi$ variable is related to the $\psi$ "radial" coordinate by $\chi=2\arctan{\exp{\psi}}$, and only the region $\psi>0$
    is shown. }
  \end{figure}

{The energy diagram of the 1-Skyrmion on \tc is shown in Fig.
\ref{fig:}(a).} {It} is divergent both for $L\to0$ and for
$L\to\infty$ and has a single minimum. This diagram resembles
qualitatively the behavior of the energy of the identity solution
$2\arctan\exp{\psi}$ (\ha{1}) on \ts rather than that of the
1-Skyrmion (\sk{1}) on \ts, of which energy is finite in the limit
of infinite radius.\footnote{{We use the same conformal
identification $\chi\equiv2\arctan{\br{\exp{\psi}}}\in(0,\pi)$ of
metric geometries of \ts and of \tc as that defined in section
\ref{sec:intro} (the spherical sections of constant $\chi$ on \ts
and of constant $\psi$ on \tc were identified).}} To understand
qualitatively why it is so, it must be remembered that \ha{1} on \ts becomes
unstable as the radius passes through its critical value
$L=\sqrt{2}$. The instability mode is associated with conformal
deformation of {\ha{1}}. Because of the instability \ha{1} bifurcates
at this radius, and \sk{1} separates from it, remaining stable for
all radii. The energy of \sk{1} tends to a finite limit, whereas
the energy of \ha{1} diverges in the limit of infinite radius. On
 \tc there is no similar bifurcation
{-- the 1-Skyrmion on \tc} is always (marginally) stable. As we
have seen, the lowest (zero) energy eigenmode is associated with
$\psi$-- translations, which preserve energy on \tc. Therefore, no
other solution  with lower energy {and the same symmetry} can
appear by a bifurcation from the 1-Skyrmion on \tc. In the limit
of infinite radius the 1-Skyrmion on \tc tends to the harmonic map
$2\arctan\Exp{\sqrt{2}\psi}$ on \tc, which is (marginally) stable
against radial and spherically symmetric
perturbations.\footnote{\label{ft:} The harmonic map on {metric}
\tc is $Y=2\arctan{\exp{\sqrt{2}\psi}}$ and has energy
$E=16\pi\sqrt{2}$. It is the critical point of the sigma model
energy functional on  \tc
$$4\pi\,\int\limits_{-\infty}^{+\infty}\ud{\psi}\br{Y'^2+2\sin^2Y}$$
 (spherical symmetry and the
hedgehog ansatz have been assumed). This map is the counterpart of
the identity solution on \ts. The same reasoning as in the text
leads to the conclusion that the map is the only finite energy
solution with nonzero topological charge and that it is
(marginally) stable against radial and spherically symmetric
perturbations.} The energy of a harmonic map must diverge as
$L\to\infty$ {because} the sigma term in the Skyrme Lagrangian
scales as $L$ and dominates the Skyrme term, which scales as
$L^{-1}$.

The minimum energy (\ref{eq:min_en}) of the 1-Skyrmion on \tc is
over $6$ times closer to the Bogomolny bound than the energy of
the 1-Skyrmion on flat space, which is
$\approx1.23145\times12\pi^2$. We remindWe must remember that the Bogomol'nyi bound
$12\pi^2$ in the \su Skyrme model is saturated on the unit
three-sphere.

The minimal energy of the approximated 1-Skyrmion profile
(\ref{eq:approx}) is only $0.37\%$ more than the true minimum
(\ref{eq:min_en}). The asymptotics of energies of the approximated
and the exact 1-Skyrmion are the same:
$$\lim\limits_{L\to0}\frac{E(L)}{\widetilde{E}(L)}=1=\lim\limits_{L\to\infty}\frac{E(L)}{\widetilde{E}(L)}.$$
Also the limiting solutions (\ref{eq:limit_sol}) are correctly
reproduced by this approximation at $L=0$ and $L=\infty$. In this
sense the profile (\ref{eq:approx}) very well approximates the
1-Skyrmion solution on \tc (\ref{eq:exact_sol}). In Fig.
\ref{fig:}(b) the exact shape function is compared with the
approximated one at $L=L_m$ [cf. Eq. (\ref{eq:min_en})].

The metric geometry of \tc is conformally identical to that of
 \ts. The manifolds may be also topologically identified if
fields with appropriate asymptotics are considered {on these
manifolds}. We have shown that (up to symmetries) only one,
topologically nontrivial {static, spherically symmetric,} and
finite energy {hedgehog} solution exist on
 \tc (it has unit topological charge). In the same topological
sector on metric \ts, arbitrarily many solutions may exist, the
number of which increases with the three-sphere's radius. The
number of possible  solutions {of this kind} is related to the
number of SU(2) harmonic maps on these manifolds. On \tc only one
harmonic map exists, whereas on \ts two countable families of
harmonic maps exist \cite{bib:biz}. Correspondingly, the structure
of solutions on \ts is very rich and the number of possible
solutions grows with $L$ \cite{bib:brat}, whereas on \tc the
structure is very simple and $L$-independent.

The structure of solutions of the Skyrme model depends on what
kind of base space is considered. It is evident that it is not the
topology of the space but its metrical properties that are important
for this structure. It is also evident that this structure is
affected by the number of harmonic maps possible on this space. It
would be therefore interesting to analyze Skyrmions on a class of
other spaces with the general line element
$\ud{\psi}^2+a^2(\psi)\br{\ud{\theta^2}+\sin^2{\theta}\ud{\phi}^2}$
[the already studied cases include $a(\psi)\propto\psi$,
$\sin{\psi}$, $\sinh{\psi}$, and at last $a(\psi)\propto1$ in this
paper] and to find out how this structure is related to the
function $a(\psi)$.

\appendix
\section{Three fundamental elliptic functions}
We used the following definitions of the standardized elliptic
integrals ($k^2<1$) \cite{bib:ryzh}:
\begin{enumerate}
\item[I]
$\qquad{}\mathbf{F}(\phi,k)=\int\limits_0^{\sin{\phi}}\frac{\ud{x}}{\sqrt{\br{1-x^2}\br{1-k^2x^2}}},$
\item[II]
$\qquad{}\mathbf{E}(\phi,k)=\int\limits_0^{\sin{\phi}}\sqrt{\frac{1-k^2x^2}{1-x^2}}\ud{x},$
\item[III]$\qquad{}\mathbf{\Pi}(\phi,\nu,k)=\int\limits_0^{\sin{\phi}}\frac{\ud{x}}{
\br{1+\nu\,x^2}\sqrt{\br{1-x^2}\br{1-k^2x^2}}}.$
\end{enumerate}
\end{document}